# Optical studies of melanin films as a material for solar light absorbers

T. Obukhova[1], M. Semenenko[2,3], M. Dusheiko[1], S. Davidenko[4], Ye. S. Davidenko[4], S. S. Davidenko[4], S. Malyuta[1], S. Shahan[2], O. S. Pylypchuk[3], P. Mochuk[1], M. Voitovych[2], T. Kuzmenko[5], A. Sarikov[2,5]

[1] Microelectronic Department, Faculty of Electronics, National Technical University of Ukraine "Igor Sikorsky Kyiv Polytechnic Institute", 16 Politehnichna Street, 03056 Kyiv, Ukraine

[2] V. Lashkaryov Institute of Semiconductor Physics of the National Academy of Sciences of Ukraine, 41 Nauky Avenue, 03028 Kyiv, Ukraine

[3] Institute of Physics of the National Academy of Sciences of Ukraine, 46 Nauky Avenue, 03028 Kyiv, Ukraine

[4] ABC Smart Solutions AG, Aegeristrasse 5, CH-6300 Zug, Switzerland

[5] Educational Scientific Institute of High Technologies, Taras Shevchenko National University of Kyiv, 4-g Hlushkova Avenue, 03022 Kyiv, Ukraine

**Abstract**

This study investigates optical and electrical properties of thin melanin films self-organized grown from water solutions of eumelanin extracted from edible plants. The melanin films exhibit the characteristics of a transparent low conductive polymer with stable in time electrical parameters. The films demonstrate indirect allowed optical transitions, an optical band gap $E_g = 1.7$ eV and a phonon energy $\hbar\omega = 0.185$ eV as well as a high absorption coefficient in the ultraviolet range, which highlight the film potential for applications as light-absorbing layers in photovoltaic solar energy converters. Analysis of infrared transmission spectra evidences presence of a significant amount of OH groups in the films, pointing to their hydrophilicity, which may have an effect on the film electrical conductivity. It is hypothesized as well that the ratio of sp³ to sp² hybridized $CH_n$ bonds influences the films' optical and electrical properties and that higher concentration of the sp² bonds may increase the conductivity due to the enhanced π-electron delocalization.

## 1. Introduction

Melanins are a family of natural pigments that are ubiquitously found in the bodies of mammals and invertebrates [1-3]. These substances provide organisms with a wide range of functionality such as pigmentation, photoprotection, free radical absorption, radiation protection, and thermoregulation. Moreover, due to their semiconductor properties, these organic composites are extremely interesting for applications in the field of electronics [1-6]. The first experimental evidence of a semiconductor behavior of melanin was reported in the early 1970s [7-11]. The surge in organic electronics in the 2000s shifted the interest toward photophysical and electronic properties of 5,6-dihydroxyindole (DHI), which is a primary structural component of eumelanin, the most common form of melanin [12]. The goal was to unravel the structure-property-function relationships of the polymers made up by DHI molecules and adapt their unique π-electron system for technological applications. Since that time, the DHI has been actively investigated as a material for electronics [13-16]. Several interesting properties of eumelanin polymers have been found out, such as nearly complete absorption of ultraviolet light [5] and mixed electronic-ionic conductivity [4, 8, 17-21]. The electric characteristics of this material such as conductivity and dielectric function were studied in [18-22] and the results of Fourier Transform Infrared (FTIR) studies are presented in [3, 6, 19, 23]. A significant number of studies considered the samples prepared by pressing melanin powder into tablets or pellets [4, 7], while in other works, melanin films obtained by dropping on a substrate with subsequent drying in different regimes were investigated [5, 8].

The melanin's broad absorption spectrum and almost complete absorption of ultraviolet radiation make this material perspective for use in photovoltaic (PV) converters – specifically as light-absorbing layers that also act as the converter's base [4, 13, 21]. To ensure this functionality, one must grow thin melanin films with sufficient conductivity values, in the range necessary for such devices. For instance, silicon-based photovoltaic converters often use substrates with the resistivity values around 1 Ω·cm (conductivity of 1 S/cm), which provide a good combination of sufficient conductivity and efficient charge carrier transport for high photoelectric conversion [24, 25]. The highest ever recorded melanin conductivity of 318 S/cm, which is significantly higher than typical values for organic semiconductors, was reported on vacuum annealed pressed melanin tablets [22]. Such high value was explained by reorganization of bonds, leading to overlap of electron density in neighboring packed chains and delocalization of their electron wave functions. Unfortunately, this study did not provide any hypotheses on how bond reorganization occurs as a result of vacuum annealing and how exactly it enhances conductivity. However, at a conductivity of 1 S/cm (i.e. a resistivity of 1 Ω·cm analogous to Si based devices), the melanin films will already be suitable for light-absorbing layers or as a base material in PV converters.

A melanin polymer is composed by a large number of oligomers, and all its properties are defined by both their relative orientations and the hydration level [17, 19, 22]. The structure of a melanin film created by oligomer stacks can change depending on various environmental factors during film polymerization [1]. Therefore, physical properties, including conductivity, of such films can be modified by film production and/or additional treatment methods.

Many studies confirm the influence of water on the melanin conductivity [17, 19, 22] showing an exponential growth in the conductivity upon increasing the hydration level. However, no mechanism has been proposed so far to explain the increase in the conductivity due to higher OH group concentrations within the films. It is known that the hydration level can be influenced by additional film treatments such as low- and moderate-temperature annealing in vacuum. Therefore, FTIR spectroscopy is appropriate to use in a study of grown melanin films to identify specific chemical bonds, especially the OH groups effecting the conductivity, in the film structure and to track bond reorganization resulting from the heat treatments.

This study examines optical and electric characteristics as well as stability of the latter of thin melanin films self-organized grown from the eumelanin extracted from edible plants. Moreover, we used FTIR spectroscopy to identify the bonding structure of the films to retrieve the information about the presence of OH groups and other complexes that may define the film conductivity.

## 2. Experimental

Melanin films were obtained by dropping a melanin water solution prepared by the method described in [26], onto a substrate with subsequent drying in air. The melanin concentrations in the solutions were 0.1, 0.5, and 0.8 % in the distilled water. Two types of substrates were used. The first substrate type, schematically shown in Fig. 1, was configured for measuring electric characteristics of the melanin films. A counter-pin Ti-Ni contact grid with $N = 132$ pins was fabricated on a ceramic (Sitall-$Al_2O_3$) plate by magnetron deposition and photolithography. Since melanin has quite high resistance, use of a counter-pin structure enables increasing the current values during electrical measurements. The pin width was 50 μm, their height was 3 μm, the distance between the pins was 50 μm, the dimensions of the grid were 6.6×5.2 mm$^2$, the length of the pathway between the pins from the top to the bottom of the structure was 34.3 cm, and the contact area of melanin films was equal to 0.01 cm$^2$.

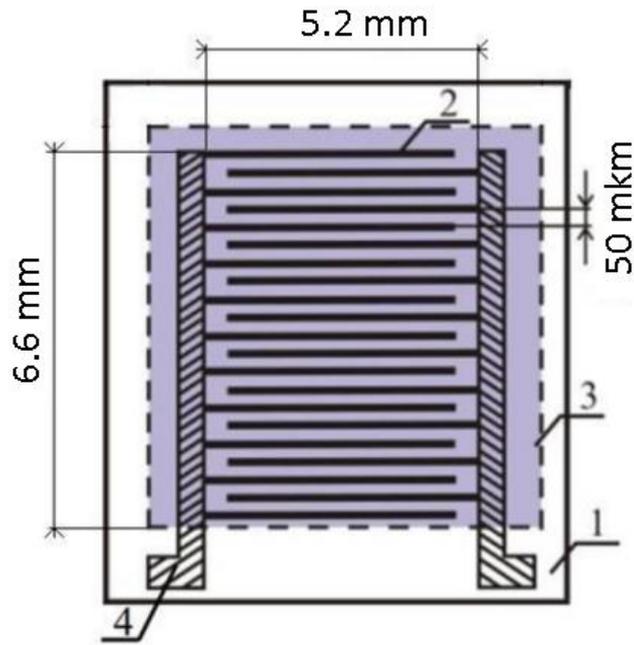

Fig. 1. Counter-pin Ti-Ni contact grid substrate: 1 – Sitall plate, 2 – pins, 3 – melanin area, 4 – metal contact grid.

The second type of substrates, namely standard microscope slide glass substrates, were used for optical and atomic force microscopy (AFM) measurements.

The AFM investigations of the surface of the prepared melanin films were carried out using a NanoScope IIIa Dimension 3000 scanning probe microscope (Bruker Inc.) in the semi-contact mode. Silicon probes TESPA-HAR (Veeco Inc.) with a nominal tip radius of 10 nm were used as the working elements.

Optical density (OD) spectra, i.e. spectral dependences of the logarithm of inverse transmission coefficient were obtained from the measurements by an A 4802 UV/VIS double beam spectrophotometer (Unico) equipped with an external Si photodiode. Namely, the photodiode voltages generated by a base beam passing through a clean glass substrate and the beam passing through a sample (a melanin film grown on a glass substrate), $U_b$ and $U_s$, were measured. The wavelength range of the base light beam was 400 to 1100 nm. The OD values were calculated as $\ln \frac{U_b}{U_s}$. From these values, spectral dependences of the absorption coefficient α were expressed as follows using the Lambert-Beer law [5]:

$$\alpha = \frac{1}{d} \ln \frac{U_b}{U_s} \qquad (1)$$

where $d$ is the melanin film thickness.

Current-voltage (*I-V*) measurements were carried out at room temperature to determine the direct current conductivity of the melanin films and to estimate the stability of their electrical characteristics. For all the samples, the *I-V* characteristics were measured on just fabricated films and repeated 4 months thereafter. The measured *I-V* dependences were recalculated into the current density versus applied field (*J-E*) dependences for convenience of analysis.

Capacitance-frequency characteristics at fixed frequency values of 100 Hz, 120 Hz, 1 kHz, 10 kHz, and 100 kHz were measured using a Uni-T UT603 meter and at 1 MHz using an E7-12 RLC meter, respectively. The devices also output the values of the tangent of the dielectric loss angle and the alternating current conductivity of the investigated samples.

Infrared transmission spectra were measured at room temperature using a PerkinElmer Spectrum BXII FTIR spectrometer. The measurements were carried out in the spectral range of 350 to 7800 $cm^{-1}$ with the resolution of 4 $cm^{-1}$ and measurement accuracy of 0.5 %. A clean glass substrate (without a melanin film) was used as a reference sample.

**3. Results and discussion**

In this work, we used atomic force microscopy to evaluate the surface homogeneity of the grown melanin films as well as to determine their thicknesses at the edge of a scratch. The AFM results for the films grown from the 0.1 and 0.5 % solutions are presented in Fig. 2. As can be seen from Fig. 2a, the melanin films obtained from the 0.1 % solution were highly inhomogeneous and had the surface roughness ranging from 0 to about 0.4 μm. Fig. 2b shows a step profile for this film, which provides the film thickness of about 0.33 μm. This value lies within the surface roughness range. Hence, the films grown from the 0.1 % solution are discontinuous and therefore will not be considered any further in this work.

The films prepared from the 0.5 % solution had a thickness of about 1.4 μm (see Fig. 2c). For a concentration of the solution of 0.8 %, the film thickness was too high to be measured by AFM. Therefore, it was determined by an interference method using a microinterferometer MII-4 to be ~ 3 μm.

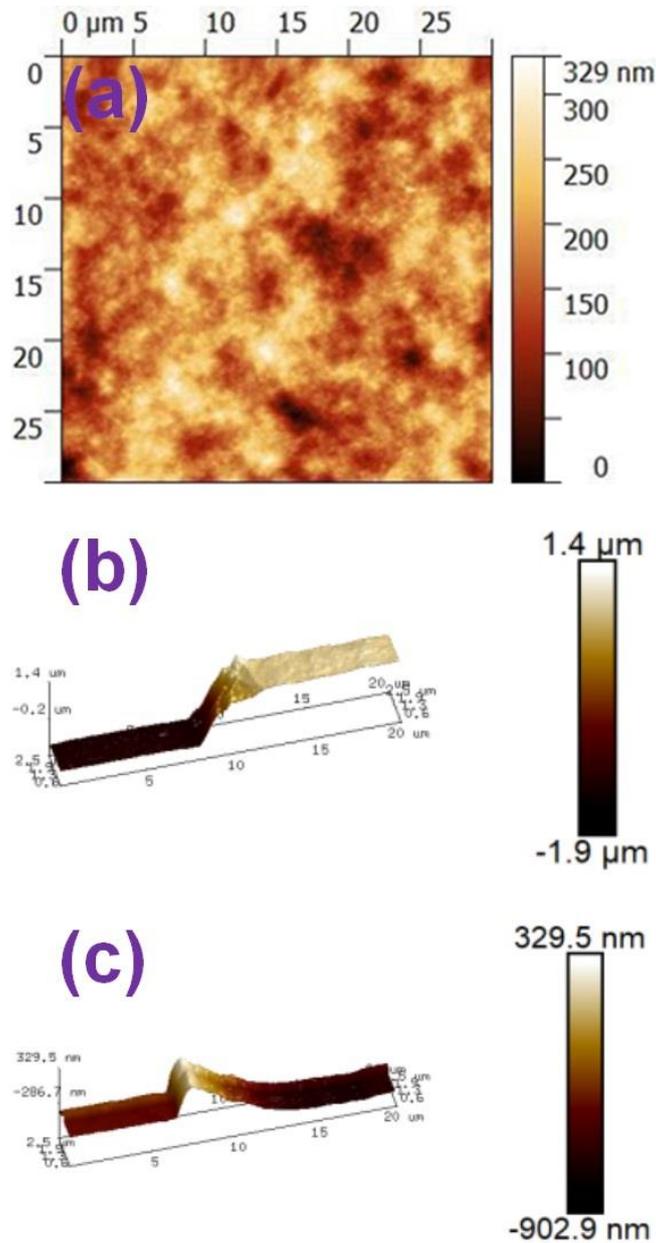

Fig. 2. (a) AFM map of surface roughness for the melanin film grown from 0.1 % solution. (b) and (c) – AFM step profiles of the melanin films grown from 0.1 and 0.5 % solutions, respectively.

The current density versus applied electric field (*J-E*) characteristics of the melanin films obtained from the 0.1 and 0.5 % solutions showed a weakly reproducible behavior. The conductivity of such films was lower as compared to that of the films prepared from the 0.8 % solution. Therefore, further results will be presented only for the latter films. The *J-E* characteristics of such film measured just after the film preparation and 4 months later are shown in Fig. 3. As can be seen from this figure, the obtained characteristics were stable in time. The difference in the current values for two measurements did not exceed 2 %. The direct current (DC) resistivity of the investigated film, $\rho = \dfrac{E}{J}$, was determined from the slope of the linear part of the *J-E* characteristics to be ρ ≈ $10^8$ Ω·cm thus giving the DC conductivity value of about $10^{-8}$ S/cm. Such result has a good agreement with the previous studies [17, 19], where highly hydrated melanin tablets were investigated.

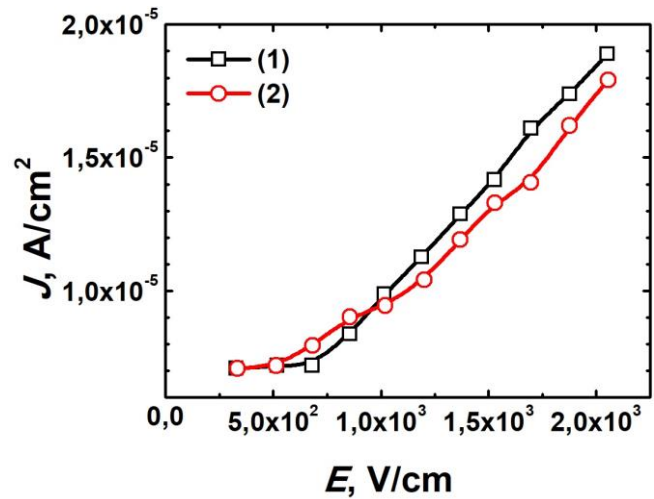

Fig. 3. *J-E* characteristics of a melanin film grown from the 0.8 % solution measured just after the film preparation (1) and 4 months thereafter (2).

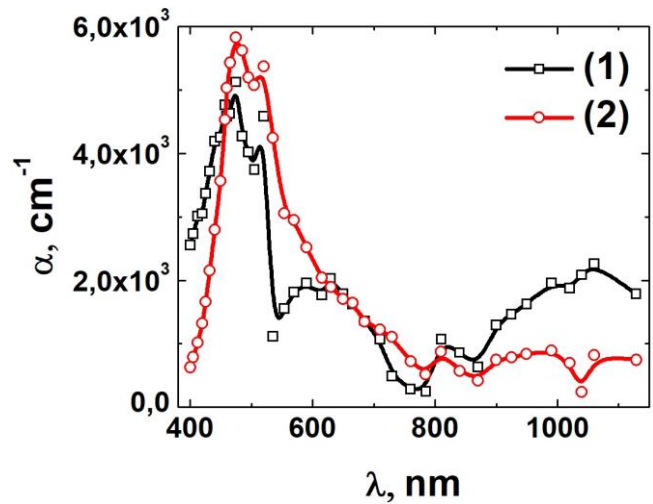

Fig. 4. Spectral dependences of the absorption coefficient of a melanin film grown from the solutions with the melanin concentration of 0.5 (1) and 0.8 (2) %.

Fig. 4 shows spectral dependences of the absorption coefficient for the melanin films prepared from the solutions with the melanin concentrations of 0.5 and 0.8 %, calculated from the respective OD spectra by Eq. (1). As can be seen from this figure, both film types have high values of absorption coefficient exceeding $6×10^4$ $cm^{-1}$ in the wavelength range of 460-500 nm corresponding to the photon energies between about 2.70 and 2.48 eV. The absorption coefficient decreases to ~ $10^4$ $cm^{-1}$ in the visible and near infrared region. Therefore, it is reasonable to consider these films as efficient light absorbers of near-ultraviolet radiation. As discussed in [5, 8], optical transitions in melanin films deposited on glass substrates are indirect. Hence, the following expression for the absorption coefficient in the case of indirect allowed optical transitions [4, 5] may be used:

$$\alpha \sim (\hbar\omega - (E_g \pm \hbar\omega_{phonon}))^2 \qquad (2)$$

where $E_g$ is the optical bandgap, $\hbar = 1.05 \times 10^{-34}$ J·s is the reduced Planck's constant, and $\omega_{phonon}$ is the frequency of phonons participating in the transitions, respectively.

Fig. 5 shows the dependence $\alpha^{1/2}(\hbar\omega)$ (Scanlon plot) recalculated from the data presented in Fig. 4. From this plot, an optical band gap value for indirect allowed transitions can be determined [27]. According to Eq. (2), the intersections of two straight lines shown in Fig. 5 with the horizontal axis correspond to the values $E_g + \hbar\omega_{phonon}$ and $E_g - \hbar\omega_{phonon}$. From this, the phonon energy and the allowed bandgap width were found to be $\hbar\omega = (1.87 - 1.5) / 2 = 0.185$ eV and $E_g = 1.5 + 0.185 = 1.7$ eV.

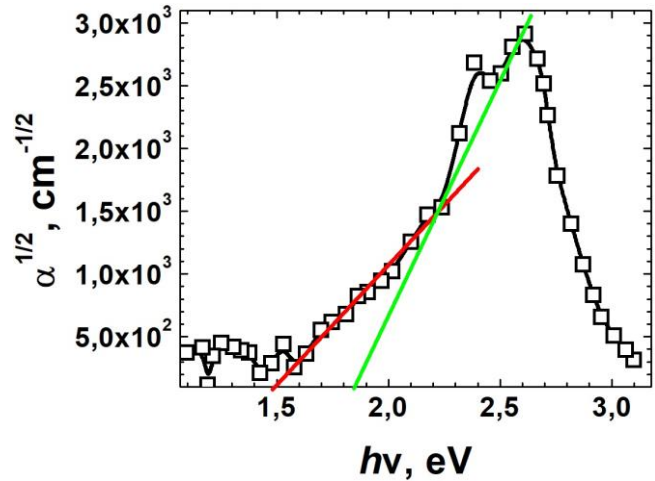

Fig. 5. Scanlon plot for the melanin film obtained from the 0.8 % solution. Straight lines define the bandgap value in the case of indirect allowed transitions.

Fig. 6 shows the measured capacitance-frequency characteristic and the frequency dependence of the tangent of the dielectric loss angle for a melanin film grown from the 0.8 % solution. From these characteristics, the conductivity at alternating current may be calculated as follows [18]:

$$\sigma = \frac{C \cdot d \cdot \text{tg}\delta}{A} \cdot 2\pi f \qquad (3)$$

where $C$ is the sample capacitance, tg δ is the tangent of the dielectric loss angle, $f$ is the applied voltage frequency during measurements, and $A$ is the sample area, respectively.

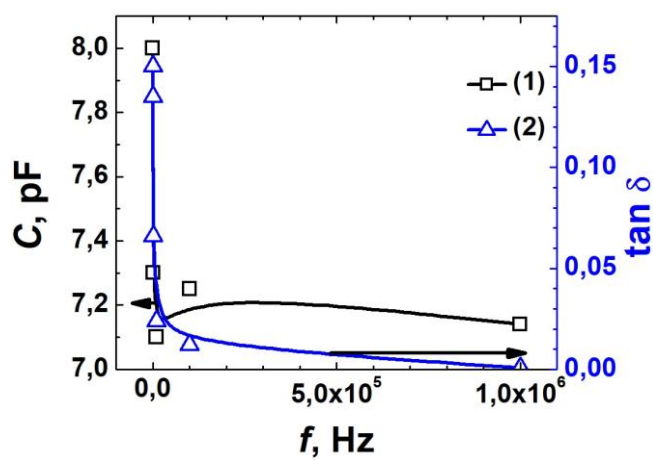

Fig. 6. Frequency dependences of the capacitance (1) and tangent of the dielectric loss angle (2) for the melanin film grown from the 0.8 % solution.

The results of calculations by expression (3) demonstrate that the alternating current conductivity of the studied melanin films does not exceed $10^{-10}$ S/cm. Measurements of the real part of the dielectric function according to [18] indicate the dielectric permittivity ($\varepsilon_{real}$) values of 2.8 at 100 Hz and 2.5 at 1 MHz, which point to absence of film doping. As discussed in [17, 19, 20, 22], the melanin conductivity is due to extensive π-conjugation, at which electrons are delocalized across the entire oligomer stack. However, melanin films also have high densities of trap states [17, 19, 20]. These traps arise from structural disorder (stack disorder), distribution of the electron energy levels within structural components of melanin oligomers or stacked polymer, and presence of radical ions. The traps induce charge carrier localization, which results in a decrease in the number of delocalized charge carriers taking part in the conduction process.

As mentioned above, the degree of melanin film hydration may have an effect on the film conductivity. An exponential growth in conductivity with increasing hydration level was reported in [17, 19, 20]. FTIR transmission spectrum of our grown eumelanin film (see Fig. 7) reveals the presence of OH groups. Therefore, the film conductivity may be indeed linked to these groups, in agreement with the findings [3, 6, 23].

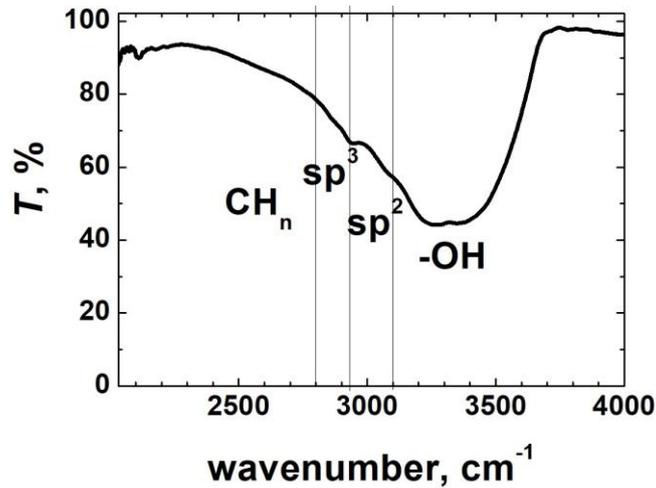

Fig. 7. FTIR transmission spectra of a melanin film grown from the 0.8 % solution.

On the other hand, the FTIR spectrum of our film also contains the bands in the range of 2800-3100 cm$^{-1}$ corresponding to the sp$^2$ and sp$^3$ hybridized CH$_n$ bonds [28, 29]. These bands are conditionally separated in Fig. 7 into the sp$^2$ and sp$^3$ related groups influencing electrical and optical properties, respectively. They may include a symmetrical CH$_2$ stretching band at 2850 cm$^{-1}$, a symmetrical CH$_3$ stretching band at 2870 cm$^{-1}$, a CH stretching band at 2915–2920 cm$^{-1}$, an asymmetrical CH$_2$ stretching band at 2926 cm$^{-1}$, and an asymmetrical CH$_3$ stretching band at 2962 cm$^{-1}$, all of them corresponding to sp$^3$ hybridized CH$_n$ bonds [28, 29]. Moreover, olefinic CH$_n$ stretching bands at 2975, 3000, 3025 and 3085 cm$^{-1}$, corresponding to sp$^2$ hybridized CH$_n$ bonds may be also observed [28, 29]. Therefore, one may hypothesize that the ratio between the sp$^2$ and sp$^3$ hybridized CH$_n$ bonds may also have an effect on the optical and electrical properties of melanin films. In particular, higher amounts of sp$^2$ hybridized bonds may be expected to increase the film conductivity [29]. In this work, a method of creating conductive channels in diamond-like carbon films by controlled electrical treatment was proposed. A reproducible increase of the film conductivity at initial sp$^2$/sp$^3$ ratios above 0.16 was observed after electrical breakdown. The possibility to vary the sp$^2$/sp$^3$ hybridized CH$_n$ bonds ratio as a tool of manipulating the conductivity of melanin films will be considered in our forthcoming publications.

**4. Conclusion**

Thin melanin films, self-organized grown from water solutions of eumelanin extracted from edible plants, exhibit the properties of a transparent, low-conductive polymer, which may be attractive for optical absorption and transportation of photogenerated carriers to other elements of a solar cell, such as the p-n junction and the positive electrode. By combining optical and electrical methods of study, we determined that the melanin films have indirect allowed transitions with the bandgap $E_g$ = 1.7 eV and the phonon energy $\hbar\omega$ = 0.185 eV. High absorption coefficient of the films in the UV range highlights the film potential for optoelectronic applications. FTIR studies revealed that the melanin films contain a significant amount of OH groups as well as $CH_n$ bonds with both $sp^3$ and $sp^2$ hybridizations. We assume that the ratio between the $sp^2$ and $sp^3$ hybridized $CH_n$ bonds may play a role on defining the film optical and electrical properties. At this, higher concentrations of $sp^2$ hybridized bonds may lead to an increase in conductivity, as these bonds likely contribute to enhanced π-electron delocalization within the polymer matrix. Therefore, the $sp^2/sp^3$ bond ratio in melanin films may be the key to optimize their performance for specific electronic and optical applications.

**Acknowledgement**

This work has been supported by the project 4Ф-2024 "Multilayer structures with organic polymer semiconductor heterojunctions and Si photon crystal based backside reflectors for photoelectric solar energy converters" of the Program of Joint Research Projects of Scientists of the Taras Shevchenko National University of Kyiv and the National Academy of Sciences of Ukraine in 2024-2025.

**Authors' contributions**

T. Obukhova – experimental setup, sample preparation, measurements of current-voltage characteristics, optical density and time stability spectra, data processing and analysis, discussion of results.

M. Semenenko – analysis and modeling of transmission spectra, analysis of capacitance-frequency characteristics, literature review, formulation of introduction and conclusions, discussion of results, formatting materials for publication.

M. Dusheiko – experimental setup, fabrication of substrates.

P. Mochuk – measurements of current-voltage characteristics and optical density spectra.

S. Davidenko, Ye. Davidenko, S. S. Davidenko – melanin production.

S. Malyuta – atomic force microscopy.

M. Voitovych – measurements and analysis of infrared transmission spectra.

S. Shahan – literature review, measurements of infrared transmission spectra.

O. S. Pylypchuk – measurements and analysis of capacitance-frequency characteristics, discussion of results.

T. Kuzmenko – analysis of transmission spectra, measurements and analysis of capacitance-frequency characteristics.

A. Sarikov – discussion of results, preparation of materials for publication.